\documentclass[preprint,nofootinbib,showpacs,aps,prc]{revtex4-1}
\usepackage{graphicx}
\usepackage{amstext}
\usepackage{amssymb}

\usepackage[usenames]{color}
\begin{document}
\title{Extracting $p\Lambda$ scattering lengths \\from heavy ion collisions}

\author{V.M. Shapoval$^a$, B. Erazmus$^{b,c}$, R. Lednicky$^{d}$  }
\author{Yu.M. Sinyukov$^{a}$}
\affiliation{$(a)$ Bogolyubov Institute for Theoretical Physics, Metrolohichna str. 14b, 03680 Kiev, Ukraine \\ $(b)$ European Organization for Nuclear Research (CERN), Geneva, Switzerland\\ $(c)$ SUBATECH, Ecole des Mines de Nantes, Universit\'e de Nantes, CNRS-IN2P3, Nantes, France \\ $(d)$ Joint Institute for Nuclear Research, Dubna, 141 980, Russia}

\begin{abstract}
The source radii, previously extracted by STAR Collaboration from the  
$p-\Lambda \oplus \bar{p}-\bar{\Lambda}$ and $\bar{p}-\Lambda \oplus p-\bar{\Lambda}$ 
correlation functions measured in 10\% most central Au+Au collisions at top RHIC energy $\sqrt{s_{NN}}=200$~GeV, differ by a factor of $2$. 
The probable reason for this is the neglect of residual correlation effect 
in the STAR analysis.
In the present paper we analyze baryon correlation functions within 
Lednick\'y and Lyuboshitz analytical model, extended
to effectively account for the residual correlation contribution.
Different analytical approximations for such a contribution are considered.
We also use the averaged source radii extracted from the hydrokinetic model (HKM) simulations 
to fit the experimental data. 
In contrast to the STAR experimental study, the calculations in HKM show both $p\Lambda$ 
and $p\bar{\Lambda}$ radii to be quite close, as expected from theoretical considerations.
Using the effective Gaussian parametrization of residual correlations
we obtain a satisfactory fit to the measured baryon-antibaryon 
correlation function with the HKM source radius value 3.28~fm.
The baryon-antibaryon spin-averaged strong interaction
scattering length is also extracted from the fit to the experimental correlation function. 

\end{abstract}

\maketitle


\section{Introduction}

The heavy ion collision experiments provide a good possibility for a study
of the baryon-(anti)baryon strong interactions using the Final State Interaction (FSI) correlation 
technique \cite{Led,fsi,FsiSin}. 
It is based on the analysis of the momentum correlations caused by final state interactions between corresponding 
baryons produced in the collision.
This activity is especially interesting in view of the ongoing nuclear collision experiments at the LHC, 
which produce great numbers of various particles, including exotic multi-strange, charmed and beauty ones. 
It allows one to study the fundamental interactions between 
specific hadron species, which can hardly be achieved by other means.
The extraction of this information makes it possible to check the correctness of
hadron-hadron strong interaction models, constrain corresponding interaction potentials,
and also improve existing cascade models (like UrQMD) by including into them the information 
about still unknown baryon-antibaryon annihilation cross-sections.

In the paper \cite{Star} the experimental $p\Lambda$ and $p\bar{\Lambda}$ correlation functions 
measured by STAR at RHIC were fitted with Lednick\'y and Lyuboshitz analytical model \cite{Led}
allowing, in principle, to extract scattering lengths characterizing the two-particle 
strong interaction. However, apart from the interaction characteristics, the correlation function 
depends also on the source spatial structure, described in terms of function $S({\bf r}^*)$,
representing the time-integrated separation distribution of particle emission points in
the pair rest frame. This fact complicates the study of the particle interaction, as it increases the 
number of free parameters which enter the fit formula. 

To simplify this study, one could calculate the corresponding source functions 
in realistic models of the collision process, which are known to describe well the experimental
observables. The hydrokinetic model \cite{HKM,Boltz,Kaon} provides 
successful simultaneous description of a wide class of bulk observables in the heavy ion collision 
experiments at RHIC and LHC \cite{Uniform}.
Moreover, it reproduces well \cite{Sf} the pion and kaon source functions for semi-central
Au+Au collisions at the top RHIC energy \cite{Phenix}, including the specific non-Gaussian tails
observed in the pair momentum and beam direction projections of the experimental source function. 
In this article we present the results of fitting the experimental data from \cite{Star}
within the analytical model \cite{Led} where the Gaussian parametrization for the 
emission source function is utilized, and the corresponding Gaussian radii are extracted from 
the HKM model simulations. 
 
\section{Models description}

The STAR collaboration studied \cite{Star} baryon-baryon $p-\Lambda \oplus \bar{p}-\bar{\Lambda}$ and
baryon-antibaryon $p-\bar{\Lambda}\oplus\bar{p}-\Lambda$ correlation functions in 10\% most central
RHIC Au+Au collisions at $\sqrt{s_{NN}}=200$~GeV. Protons and antiprotons in transverse momentum 
range $0.4<p_T<1.1$~GeV/c with the rapidity $|y|<0.5$, and lambdas and antilambdas with 
$0.3<p_T<2.0$~GeV/c and $|y|<1.5$ were selected for the analysis.

The experimental correlation function is constructed as the ratio of the distribution of 
particle momentum in the pair rest frame, $k^{*}$, in the same events to the analogous distribution 
in mixed events.
Then the measured correlation function $C_{meas}$ is corrected for the pair purity, 
defined as the fraction of correctly identified primary particle pairs among all the selected ones,
to give the corrected function $C_{corr}$
\begin{equation}\label{l2}
C_{corr}(k^{*}) = \frac{C_{meas}(k^{*})-1}{\lambda(k^{*})}+1,
\end{equation}
where $\lambda(k^{*})$ is the pair purity. The estimated mean pair purity in the experiment is $\lambda = 17.5 \pm 2.5 \%$.

To fit the experimental correlation function the Lednick\'y and Lyuboshitz analytical model \cite{Led} is used, which connects the two-particle correlation function $C(k^{*})$ with the particle emission source size $r_0 $ and the s-wave strong interaction scattering amplitudes $f^S(k^{*})$ at a given total pair spin $S$.
In the equal-time approximation, valid on condition $|t_1^*-t_2^*| \ll m_{2,1}r^{*2}$ for $\mathrm{sign}(t_1^*-t_2^*)=\pm 1$ respectively,
the correlation function can be calculated as a square of the wave function $\Psi^S_{- \textbf{k}^{*}}$, representing the stationary solution of the scattering problem with the opposite sign of the vector $\textbf{k}^{*}$, averaged 
over the total spin $S$ and the distribution of the relative distances $S(\textbf{r}^{*})$:
\begin{equation}\label{l3}
C(k^{*}) = \left\langle \left| \Psi^{S}_{- \textbf{k}^{*}}(\textbf{r}^{*})\right|^{ 2}\right\rangle. 
\end{equation}
In typical nuclear collisions the source radius can be considered much larger than
the range of the strong interaction potential, so $\Psi^S_{- \textbf{k}^{*}}$ at small $k^*$ can be approximated by the
s-wave solution in the outer region: 
\begin{equation}\label{l4}
\Psi_{- \textbf{k}^{*}}^{S} (\textbf{r}^{*}) = e^{-i\textbf{k}^{*} \cdot \textbf{r}^{*}}
+ \frac{f^{S}(k^{*})}{r^{*}} e^{ik^{*} \cdot r^{*}}.
\end{equation}
The effective range approximation for the s-wave scattering amplitude is utilized
\begin{equation}\label{l5}
f^{S}(k^{*}) = \left( \frac{1}{f^{S}_{0}} + \frac{1}{2} d^{S}_{0} k^{*2} - i k^{*} \right)^{-1},
\end{equation}
where $f^{S}_{0}$ is the scattering length and $d^{S}_{0}$ is the effective radius for a given 
total spin $S=1$ or $S=0$.

The particles are assumed to be emitted unpolarized (i.e. with the polarization $P=0$), so that the fraction of pairs 
in the singlet state $\rho_0 = 1/4 (1-P^2)=1/4$, and in the triplet state $\rho_1 = 1/4 (3+P^2)=3/4$.

The normalized separation distribution (source function) $S({\bf r}^*)= N^{-1} d^3N/d^3{\bf r}^*$
is assumed to be Gaussian one
\begin{equation}\label{l6}
S({\bf r}^*) = (2\sqrt{\pi}r_0)^{-3} e^{-\frac{\textbf{r}^{*2}}{4r_0^2}},
\end{equation}
where $r_0$ is considered as the effective radius of the source.

Under such assumptions the correlation function can be calculated analytically \cite{{Led}}: 
\begin{eqnarray}\label{LL}
&& C(k^*) = 1 + \sum_S \rho_S\left[\frac12\left|\frac{f^S(k^*)}{r_0}
\right|^2\left(1-\frac{d_0^S}{2\sqrt\pi r_0}\right)\right. + \nonumber\\
&&\left.\frac{2\Re f^S(k^*)}{\sqrt\pi r_0}F_1(2k^* r_0)-
\frac{\Im f^S(k^*)}{r_0}F_2(2k^* r_0)\right],
\end{eqnarray}
where $F_1(z) = \int_0^z dx e^{x^2 - z^2}/z$ and $F_2(z) = (1-e^{-z^2})/z$.
The term $-\frac{d_0^S}{2\sqrt\pi r_0}$ in this expression corresponds to 
the correction accounting for a deviation of $\Psi^S_{- \textbf{k}^{*}}$ from the true wave function
inside the range of the strong interaction potential. 
So, the model has quite a large number of parameters, being the scattering lengths $f^{S}_{0}$
and effective radii $d^{S}_{0}$, which may both be complex in general case, 
and the source radius $r_0$. Although in principle all of them can be determined from the measured data,
in each concrete situation the number of free parameters can be reduced by making certain reasonable assumptions about the values of some of them.

In our study the source radius $r_0$ is extracted from the Gaussian fit to the source functions calculated in 
hybrid HKM model.
The simulation of the full process of evolution of the system formed in nuclear or particle collision
in hybrid HKM consists of two stages. The first one is hydrodynamical expansion of thermally and chemically equilibrated matter described within ideal hydrodynamics approximation with the lattice-QCD inspired equation of state~\cite{Laine} (corrected for small but nonzero chemical potentials), which is matched
with the hadron-resonance gas in chemical equilibrium via cross-over type transition.
The second stage consists in gradual system decoupling after loosing chemical and thermal equilibrium.
It can be described either within hydrokinetic approach with switching to UrQMD cascade at some space-like hypersurface situated behind the hadronization one, or with sudden switch to UrQMD cascade at the hadronization hypersurface. In current study we choose the second variant of switching to cascade, basing on \cite{Uniform}, where the comparison of one- and two-particle spectra, calculated at both types of matching hydro and cascade stages for RHIC and LHC energies, showed a fairly small difference between them.

The model provides particle distribution functions $\frac{d^6N}{d^3x d^3p}$ at the chosen switching hypersurface. 
Using the Monte-Carlo procedure, one generates particle momenta and coordinates according to these distributions,
which serve as the input for the UrQMD hadronic cascade. 

To perform a specific calculation one should specify the initial conditions for the hydrodynamics stage attributed to the starting proper time $\tau_0$. These conditions are the initial energy density (or entropy) profile $\epsilon(\textbf{r})$ and the initial rapidity profile (initial flow) $y (\textbf{r})$.
Here we suppose longitudinal boost-invariance and use $\epsilon(\textbf{r}_T)$ corresponding to the MC-Glauber model  calculated
with GLISSANDO code \cite{Gliss}. The maximal energy density $\epsilon_0$ is chosen to
reproduce the experimental mean charged particle multiplicity,
and the initial flow is supposed to be $y_T = \alpha \frac{r_T}{R^2(\phi)}$, with $\alpha=0.45$~fm for top RHIC energy.  Thus, model has only two free parameters $\epsilon_0$ and $\alpha$.
We start the hydrodynamic evolution at $\tau_0 = 0.1$~fm/c. 
Sudden switch from hydrodynamics to UrQMD is performed at the isotherm $T=165$~MeV.
The hadron distribution functions (for each hadron sort~$i$) at the switching hypersurface $\sigma_{sw}$ are calculated according to the Cooper-Frye formula
\begin{equation}\label{l8}
p_0 \frac{d^3 N_i}{p_T dp_T d\phi_p dy} = \int_{\sigma_{sw}} p^\mu d \sigma_\mu f_i^{eq}(p \cdot u(x), T(x), \mu_i(x)).
\end{equation}
The procedure of filling the histograms for source functions $S(\textbf{r}^{*})$
(here $\textbf{r}^{*}$ is the particle spatial separation in the pair rest frame) 
can be described by the formula
\begin{equation}\label{l9}
S({\bf r}^{*(k)})= \frac{\sum_{n=1}^{N_{\rm ev}}\sum_{i_1^n,i_2^n}
    \prod_{\alpha=1}^3[\delta_{\Delta_\alpha}
    (r_\alpha^{*(k)}-r^*_{i_1^n\alpha}+r^*_{i_2^n\alpha})/\Delta_{\alpha}]}{\sum_{n=1}^{N_{\rm ev}}\sum_{i_1^n,i_2^n}1}
\end{equation}
Here ${\bf r}_{i_1^n}^*$ and ${\bf r}_{i_2^n}^*$ are the pair rest frame three-coordinates 
of particles 1 and 2 produced in the $n-$th event, ${\bf r}^{*(k)}$ is the three-coordinate 
of the center of the $k$-th histogram bin, the function $\delta_\Delta(x)=1$ 
if $|x|<\Delta/2$ and $0$ otherwise, and $\Delta_\alpha$ is the size of the 
$\alpha$-projection of the histogram bin.

\section{Results and discussion}

\begin{figure}[t]
\centering
  \includegraphics[bb=0 0 567 384,width=0.9\textwidth]{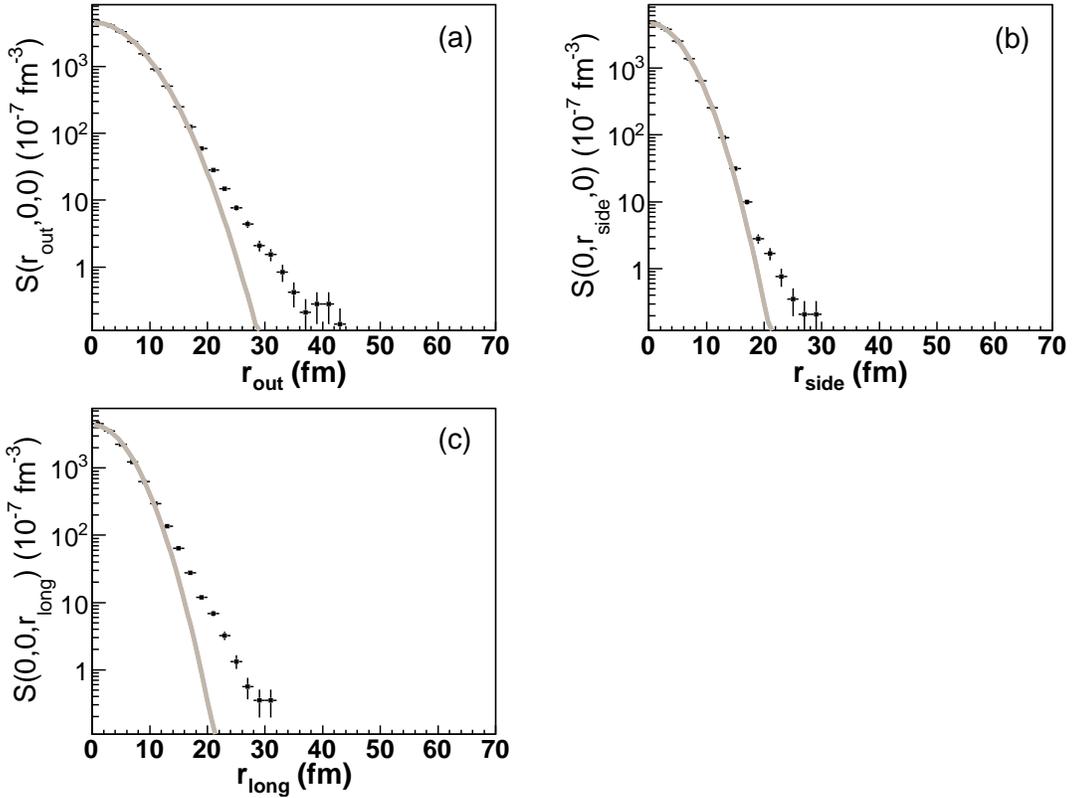}
  \caption{The $p\Lambda$ source function projections calculated in HKM (markers) and the Gaussian fits to them (lines).
The simulations correspond to 10$\%$ most central collisions at $\sqrt{s_{NN}}=200$ GeV 
in conditions of STAR experiment at RHIC \cite{Star}. Pair transverse momentum and 
rapidity cuts correspond to those in the experiment. The cut $k^{*}<50$~MeV/c 
is also applied to select the pairs from correlation effect domain, 
where the $r^*-k^*$ correlations are small.
  }\label{fig1}
\end{figure}

The $p\Lambda$ source function projections calculated in HKM together with the corresponding Gaussian fits are presented 
in Fig.~\ref{fig1}. Here the out-side-long coordinate system is used, where the \textit{out} 
axis is directed along the pair total momentum in longitudinally co-moving system,
the \textit{long} direction coincides with the beam axis, and the \textit{side} axis is 
perpendicular to the latter two ones.

\begin{figure}[t]
\centering
  \includegraphics[bb=0 0 567 384,width=0.8\textwidth]{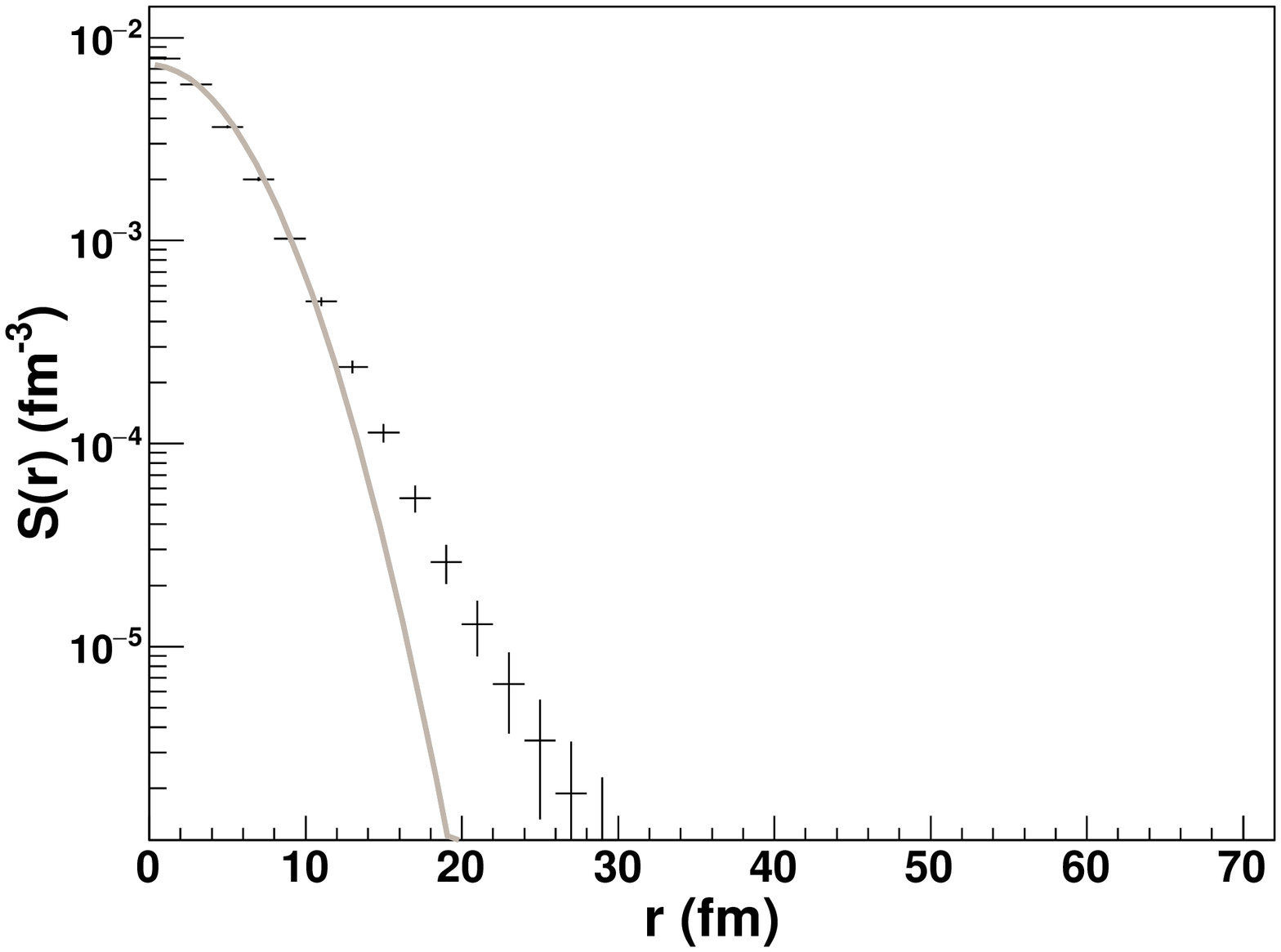}
  \caption{The $p\Lambda$ angle-averaged source function calculated in HKM (markers) and the Gaussian fit to it (line). The simulation conditions are the same as given in caption of Fig.~\ref{fig1}
    }\label{fig11}
\end{figure}

One can notice that HKM predicts $p\Lambda$ source function having non-Gaussian tails.
The similar behavior was observed experimentally \cite{Phenix} and reproduced in HKM
simulations \cite{Sf} for pion source function.
In $p\Lambda$ case these tails appear partially because of the averaging over a wide $p_T$ interval.
We also see that in different directions the corresponding Gaussian fits have
different widths (especially the \textit{out}-projection is much wider than the other ones).
These peculiarities of the HKM separation distribution require, in principle, the generalization of the existing Lednicky-Lyuboshitz 
formula~(\ref{LL}) to the cases of anisotropic non-Gaussian source functions and heavy tails ~\footnote{Note, however, that at such a model extension the number of its 
free parameters will increase significantly, and utilization of the generalized formula for the reliable description of experimental 
data will require putting additional constraints on these parameters.}.
However, for simplicity and since in the described STAR experiment 
only one-dimensional correlation function is analyzed with no respect to spatial directions, 
in current study we are going to stay within the isotropic Gaussian approximation (\ref{LL})
and for this aim utilize the angle-averaged $r_0$ value. 
It can be extracted from the Gaussian fit
to the angle-averaged source function (see Fig.~\ref{fig11})
$S(r)=1/(4\pi)\int_0^{2\pi} \int_0^{\pi} S(r,\theta,\phi) \sin\theta d\theta d\phi$.

\begin{figure}[t]
\centering
\includegraphics[bb=0 0 567 384, width=0.9\textwidth]{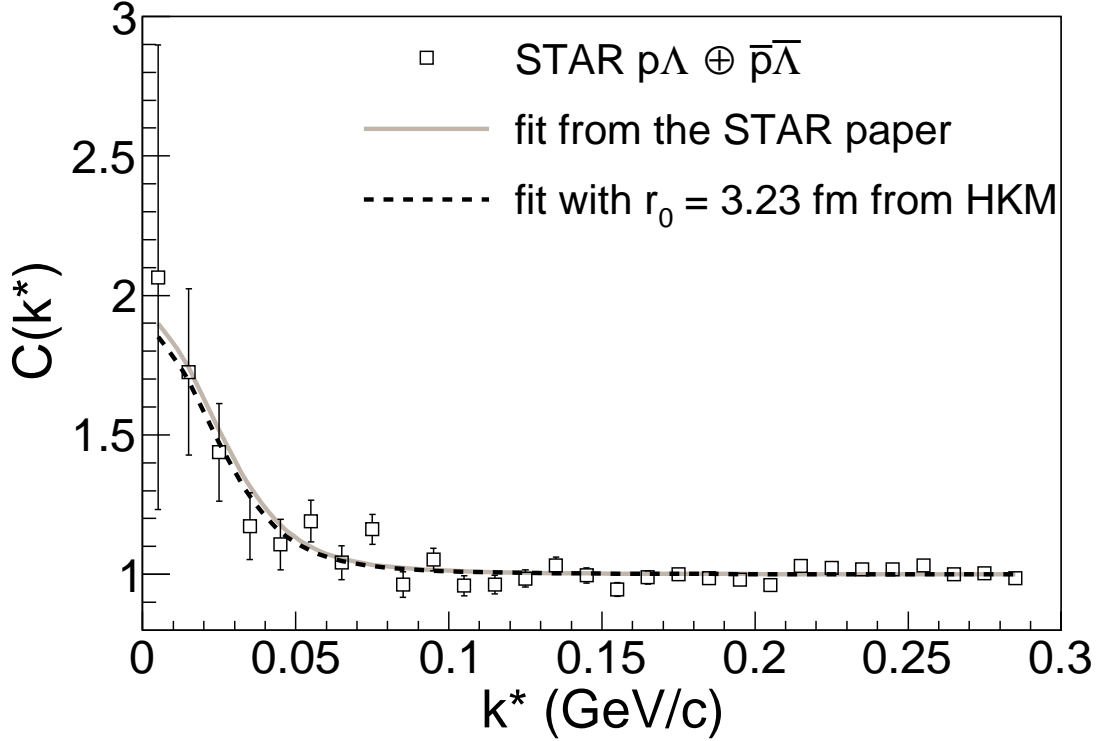}
\caption{The $p-\Lambda \oplus \bar{p}-\bar{\Lambda}$ correlation function measured by STAR (open markers),
the corresponding fit according to (\ref{LL}) with parameters fixed as in the STAR paper \cite{Star} within the Lednick\'y and Lyuboshitz analytical
model \cite{Led} (light solid line) and our fit within the same model with the source radius $r_0$ extracted from the
HKM calculations (dark dashed line).
} \label{fig2}
\end{figure}

\begin{figure}[t]
\centering
\includegraphics[bb=0 0 567 384 ,width=0.9\textwidth]{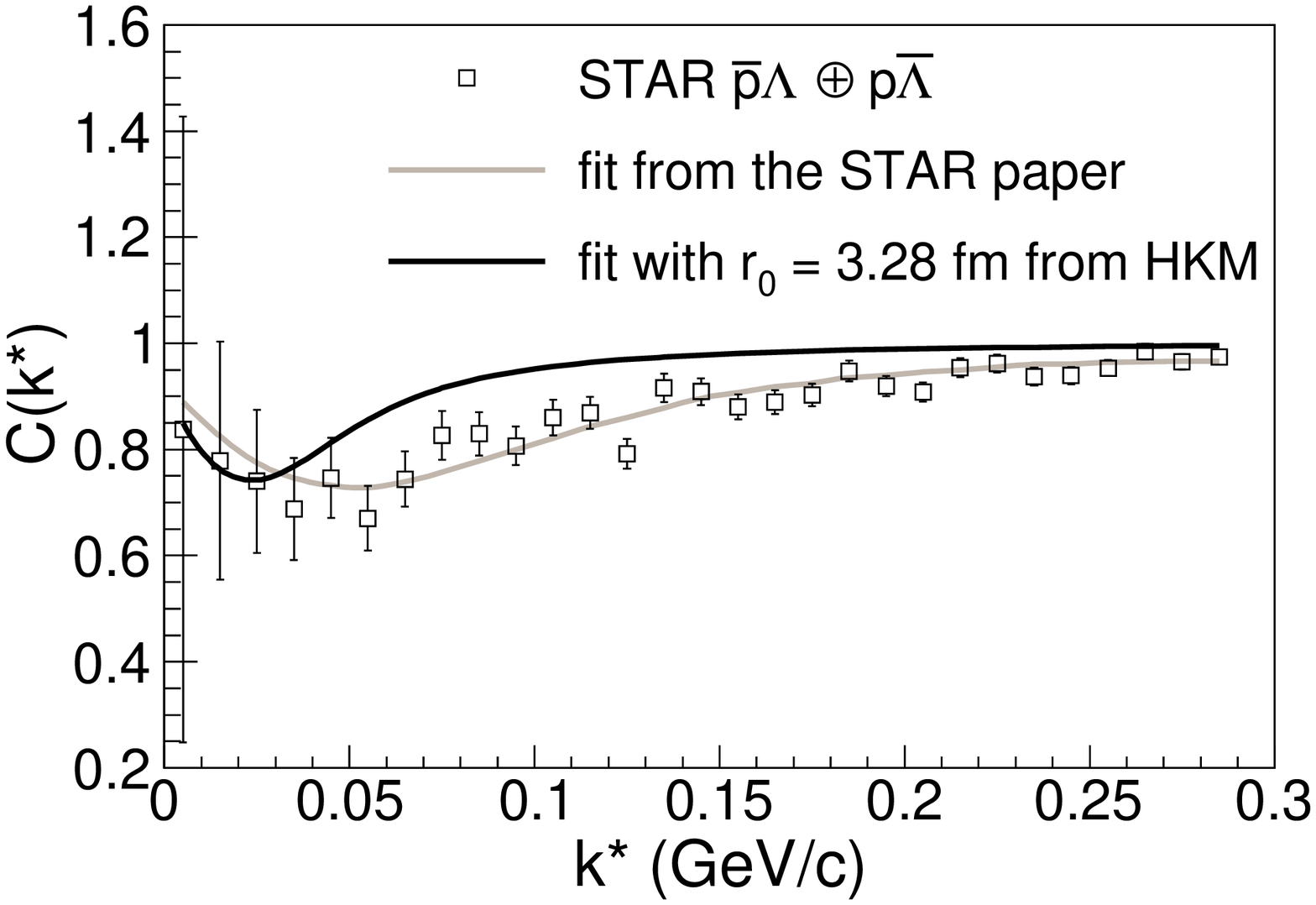}
\caption{The same as in Fig.~\ref{fig2} for the $\bar{p}-\Lambda \oplus p-\bar{\Lambda}$ correlation function.
} \label{fig3}
\end{figure}

\begin{figure}
\centering
\includegraphics[bb=0 0 567 409,width=0.9\textwidth]{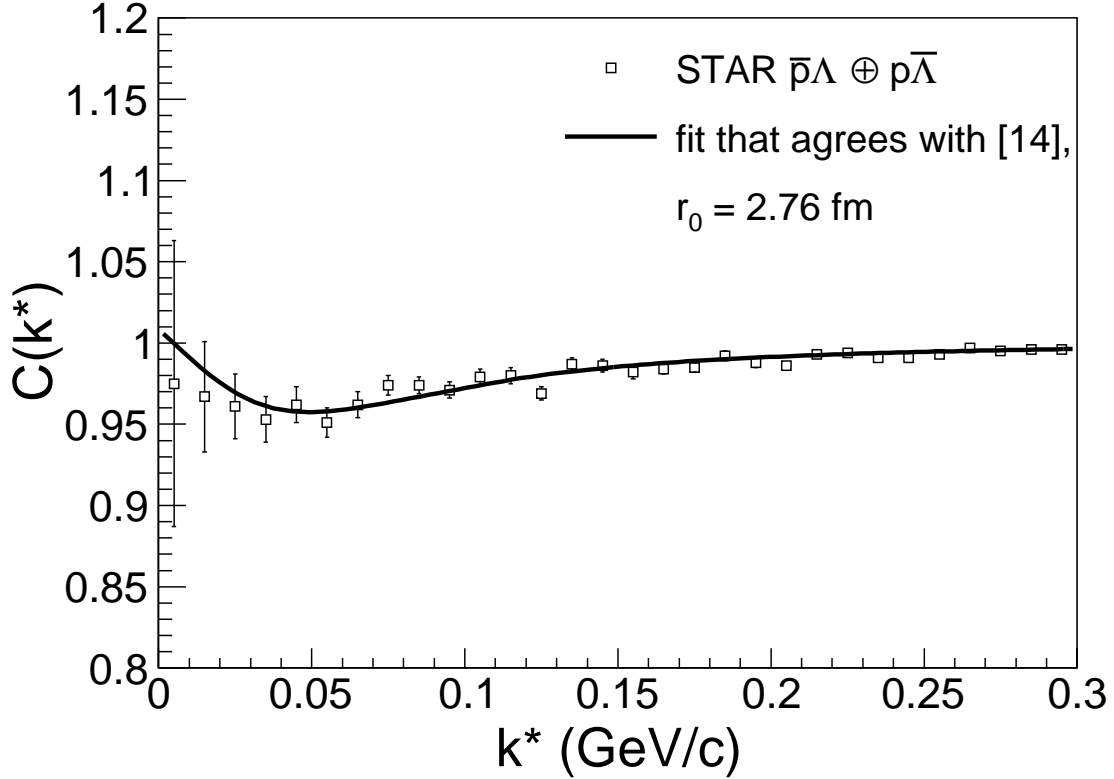}
  \caption{Our fit (black line) to the STAR purity uncorrected $\bar{p}-\Lambda \oplus p-\bar{\Lambda}$ correlation function as presented in~\cite{Kisiel} (open markers) according to (\ref{l12}), (\ref{LL}), with $C_{res}(k^*)$ term in the form (\ref{G0}). 
In this particular fitting the source radius $r_0$ is a free parameter.
The extracted fit parameter values are $r_0 = 2.76 \pm 0.13$~fm, $\Re f_0 = 0.59\pm 0.19$~fm, $\Im f_0 = 0.85\pm 0.14$~fm, with $\chi^2/\mathrm{ndf}=1.37$.
}\label{fig_kisiel}
\end{figure}

\begin{figure}
\centering
\includegraphics[bb=0 0 567 412,width=0.9\textwidth]{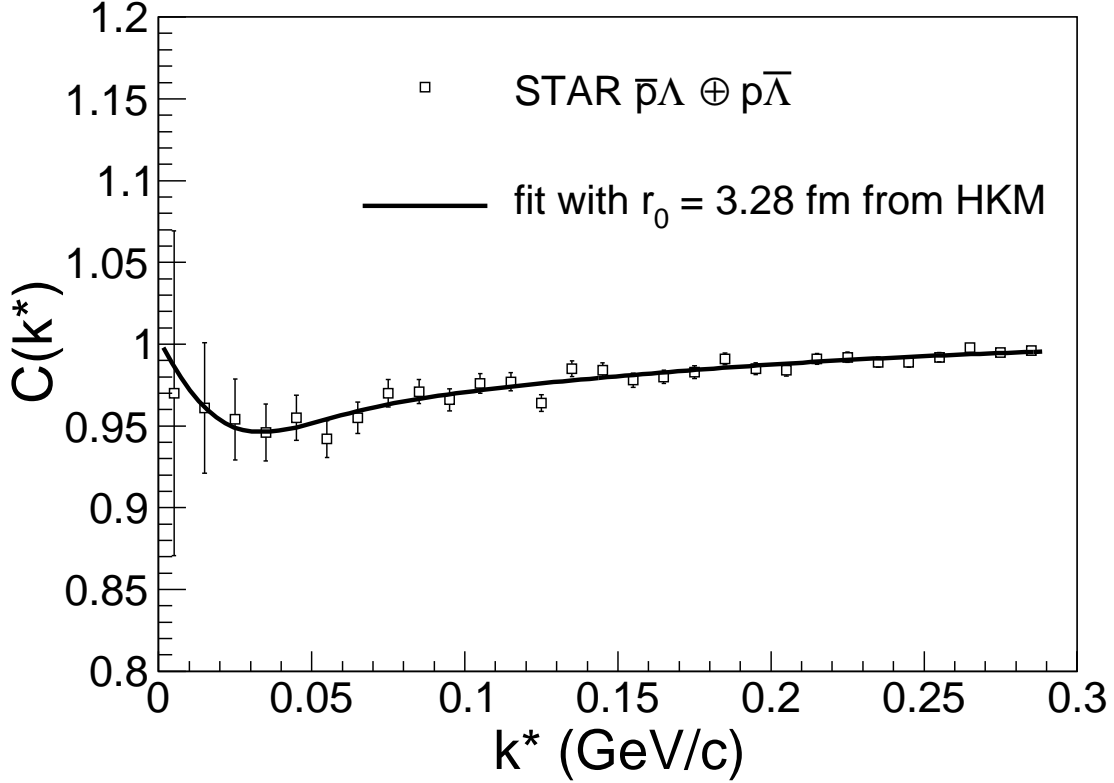}
  \caption{The purity uncorrected $\bar{p}-\Lambda \oplus p-\bar{\Lambda}$ correlation function 
  measured by STAR \cite{Renault} (open markers) and our fit to it according to (\ref{l12}) and (\ref{LL}) 
  (black line), with the Gaussian parametrization~(\ref{G}) for the residual correlation term $C_{res}(k^*)$.
The source radius $r_0$ was fixed at a value extracted from the HKM calculations.
The extracted fit parameters are $\Re f_0= 0.14\pm 0.66$~fm, $\Im f_0= 1.53\pm 1.31$~fm, $\beta=0.034\pm 0.005$ and $R= 0.48\pm 0.05$~fm, with $\chi^2/\mathrm{ndf}=0.87$.
}\label{fig4}
\end{figure}

In Figs.~\ref{fig2}--\ref{fig4} we present experimental $p-\Lambda \oplus \bar{p}-
\bar{\Lambda}$ and 
$\bar{p}-\Lambda \oplus p-\bar{\Lambda}$ correlation functions,
measured by STAR collaboration in 10\% most central Au+Au collisions at top RHIC energy $
\sqrt{s_{NN}}=200$~GeV together with the fits
performed within Lednick\'y and Lyuboshitz analytical model.  

For baryon-baryon case (Fig.~\ref{fig2}) the scattering lengths $f^{S}_{0}$ and effective radii $d^{S}_{0}$
values are taken from \cite{Wang} ($f_0^s = 2.88$~fm, $f_0^t = 1.66$~fm, $d_0^s = 2.92$~fm, $d_0^t = 3.78$~fm). Thus, the source radius $r_0$ is the only free parameter in the STAR fit~\cite{Star} (light curve). Fitting gives $r_0^{exp}=3.09 \pm 0.30_{- 0.25}^{+0.17} \pm 0.2$~fm.
In our own fit (dark curve) all the parameters are fixed, and 
$r_0^{\mathrm{HKM}}=3.23$~fm is determined from a Gaussian fit (\ref{l6}) to calculated in HKM $\textbf{r}^{*}$-distribution in the pair rest frame for $r^{*} < 70$~fm. 
One can see that HKM model radius is consistent with that extracted from the STAR data in~\cite{Star}.


In the baryon-antibaryon case (Fig.~\ref{fig3}) to reduce the number of free parameters 
both singlet and triplet scattering amplitudes are assumed to be equal, $f^s=f^t=f$ (approximately corresponding to spin-averaged scattering length $f_0$),  and both effective radii are set to zero $d_0^s=d_0^t=0$ in the STAR fit.
The scattering length should have a positive imaginary part $\Im f_0 > 0$ describing the contribution of annihilation channels and leading to a wide dip in the correlation function at intermediate $k^*$-values due to the last term in Eq. (\ref{LL}).
Thus, the model has three free parameters $\Re f_0$, $\Im f_0$ and $r_0$ in \cite{Star} and two free parameters $\Re f_0$, $\Im f_0$ in our fit.
The STAR has obtained $p\bar{\Lambda}$ source radius value $r_0=1.50 \pm 0.05^{+0.10}_{- 0.12} \pm 0.3$~fm (light curve),
which is $\sim 2$ times smaller than the $p\Lambda$ one, although there is no apparent physical reason for such a difference. Both radii can be expected to have similar values, and the HKM source radius for the baryon-antibaryon case $r_0^{\mathrm{HKM}}=3.28$~fm is expectedly close to the corresponding 
baryon-baryon one. But at this source radius the fitting curve (dark) is too narrow to describe the data points.

However, residual correlations were not taken into account in the STAR analysis~\cite{Star,Kisiel}.
Constructing the experimental correlation function one usually supposes that
only the pairs composed of two primary particles are correlated, and the rest of the pairs, which include
secondary or misidentified particles, are supposed to be uncorrelated. 
However, among such pairs the so-called residual correlations can exist. They occur if the parent of a particle from such a pair
was correlated with another particle forming the pair.
For example, if $\Sigma^{+}$, correlated with some $\Lambda$, decays into proton and 
$\pi^0$,
then this proton will be residually correlated with the $\Lambda$.
The interactions in most of such pairs are unknown, so at the moment there is no possibility to 
reliably refine the constructed experimental correlation function from the effect of residual 
correlations. However, one can try to account for the residual correlations at least 
phenomenologically using a simple analytical approximation to the residual correlation function.

Note that the effect of the residual correlations has presumably only a minor influence on 
the baryon-baryon $p\Lambda$ correlation function since in this case there are not so many
inelastic channels open for parent FSI near threshold
and if open, they are usually suppressed also being near
threshold. So, $\Im f_0$ for parent pairs (giving negative contribution
to residual correlation function) is expected to be small and the sign of $\Re f_0$
may vary from one parent pair to another, thus likely reducing the net effect 
of $\Re f_0$ on residual correlation function.

In contrast to this, in case of $\bar{p}\Lambda$ correlation function 
there is a number of annihilation channels significantly contributing to
$\Im f_0$ and leading to a substantial negative contribution to
residual correlation function.

In the case when the measured baryon-antibaryon correlation function is not corrected for purity, the fitted uncorrected correlation function is expressed through the true one in (\ref{LL}) similar to (\ref{l2}):
\begin{equation}\label{l10}
C_{uncorr}(k^{*}) = \lambda(k^{*}) C(k^{*}) + (1-\lambda(k^{*})),
\end{equation}
The pair purity $\lambda(k^{*})$ in our calculations is extracted 
according to (\ref{l10}) from the plots of $C_{\rm uncorr}(k^*)$ and $C(k^*)$ 
provided in \cite{Renault}.

The first term in formula (\ref{l10}) corresponds to the pairs of correlated (primary only) particles,
and the second one represents the contribution of the uncorrelated pairs, where one or both particles are   misidentified or secondary ones.
Assuming that among the latter there can be residually correlated pairs, one should modify this 
expression to account for the residual correlations as well.

In the recent paper \cite{Kisiel}, the account for residual correlations is performed by summarizing all the contributions from
different parent pairs to the full correlation function, making, however, a 
number of simplifying assumptions. Particularly, the purity $k^*$-dependence is neglected
in their analysis. As for the scattering parameters $f_{0i}$ and $d_{0i}$ of the parent systems
and the corresponding parent source radii $r_{0i}$, the effective range parameters $d_{0i}$ are
neglected and $f_{0i}$ and $r_{0i}$ are assumed to be equal to universal baryon-antibaryon values.

As compared with \cite{Kisiel}, we propose here the alternative approach 
avoiding such a detailed (containing however a number of assumptions) 
calculation of the residual correlations. Instead, we are aiming to describe them 
by introducing some effective residual correlation function $C_{\rm res}(k^*)$
for a fraction $\alpha(k^*)$ of the pairs of particles supposed earlier in Eq. (\ref{l10})
uncorrelated. Then,
\begin{equation}\label{l12}
C_{uncorr}(k^{*}) = 1 + \lambda(k^{*}) (C(k^{*})-1) + \alpha(k^*) (C_{res}(k^*)-1). 
\end{equation}

Obviously, Eq. (\ref{l12}) reduces to Eq. (\ref{l10}) if either
    $C_{\rm res}(k^*)=1$ or $\alpha(k^*)=0$.
    One may approximate $\alpha(k^*)$ with a constant
    (e.g., put $\alpha=0.49$, which is the total fraction
    of the pairs containing daughter particles as given in
    table III of \cite{Star}) or assume it proportional to the fraction
    of non-primary particles:
    $\alpha(k^*)=\tilde{\alpha}[1-\lambda(k^*)]$,
    where $\tilde{\alpha}$ is a fit parameter.

    Choosing the concrete form of $C_{\rm res}(k^*)$ requires
    some additional analysis. First of all, one should take into
    account that due to the phase space suppression of small
    $k^*$-values, the baryon-antibaryon correlations are dominated
    by the effect of wide annihilation dips in parent correlation
    functions related with $\Im f_{0i}$ through the negative last
    term in Eq.~(\ref{LL}). The parent decays widen these dips and
    wash out possible structures at small $k^*$ related with
    $\Re f_{0i}$. Further, the contribution to the $C_{\rm res}(k^*)$
    from a given parent correlation function recovers the latter
    for $k^*$ larger than the parent decay momenta, i.e. for
    sufficiently large $z=2k^*r_0$ \cite{Stavinsky}.
    Moreover, following from large-$z$ behavior of the functions
    $F_1(z) \rightarrow (2z)^{-2}$ and $F_2(z) \rightarrow z^{-1}$,
    the parent correlation functions approach unity from below
    according to a universal inverse power law $\propto z^{-n}$,
    where the power $n > 2$ increases with the number of terms
    essentially contributing in the effective range expansion
    in a given $k^*$-interval. Particularly, $n=3$ if one may neglect
    already the effective range parameters $d_{0i}$ (i.e. neglect
    the $k^*$-dependence of the effective range function):
    \begin{equation}
    \label{large-z}
      C_i(z)-1 \rightarrow \frac{-u+v/(\sqrt{\pi}z)}
               {z[(1+u z/2)^2+(v z/2)^2]} \rightarrow z^{-3},
    \end{equation}
    where $u=\Im f_{0i}/r_{0i}$, $v=\Re f_{0i}/r_{0i}$.
    For practical calculations, one may follow \cite{Kisiel} and assume
    approximately the same source radii and scattering parameters
    for all baryon-antibaryon systems.
    Then, one can approximate $C_{\rm res}(z)$ at large
    enough $z$ by the $\bar{p}\Lambda$ correlation function
    $C(z)$ and approximately account for the washing out effect
    of parent decays by smoothly tailing the latter at some value
    $z=z_t$ to a slowly varying function $1 - A + B z^c$:
\begin{eqnarray}\label{G0}
&&C_{res}(z, z \leq z_t) = 1 - A + Bz^c, \nonumber\\
&&C_{res}(z, z > z_t) = C(z).
\end{eqnarray}

Using the effective expression (\ref{G0}) for $C_{res}$ and data points from \cite{Kisiel},
one can reproduce the
results obtained in \cite{Kisiel} but in much simpler way. So,
    assuming $\alpha(k^*)=0.49$ and $\lambda(k^*)=0.15$ as
    in \cite{Kisiel} and fixing $z_t=3$, $c=1.5$ based on the analysis
    of various contributions to the residual correlation function
    including those in Fig.~4 in \cite{Kisiel}, the fit results are: 
$r_0 = 2.76 \pm 0.13$~fm, $\Re f_0 = 0.59\pm 0.19$~fm, $\Im f_0 = 0.85\pm 0.14$~fm,
with $\chi^2/\mathrm{ndf}=1.37$ (see Fig.~\ref{fig_kisiel}). 
They agree within the errors
with the result from \cite{Kisiel}:
$r_0 = 2.83\pm 0.12$~fm, $\Re f_0= 0.49\pm 0.21$~fm, $\Im f_0 = 1.00\pm 0.21$~fm.
The fitted $r_0$ value agrees with the fit result for
    $p\Lambda$ system, however, it is substantially smaller than the HKM prediction of 3.28 fm.
        As a result, the fit with the radius fixed at the HKM value
        leads to unsatisfactory description of the correlation function.

    Therefore, sticking on the HKM result, one is enforced
    to  apply a more flexible
     parametri\-zation of $C_{\rm res}$, avoiding the assumption
     of the universal form of baryon-antibaryon correlation
     functions at $z>z_t$ with a constant effective range function
     (i.~e. with neglected $d_{0}$ and higher order expansion
     parameters). It is worth noting that even if keeping the universality
     assumption, the account of additional complex expansion
     parameters would make the fit quite unstable and unpractical
     at given statistical errors. Instead, one can use the effective
     Gaussian parametrization with reasonable behavior at small
     and large $k^*$-values~\cite{preprint, StarPRL}:
\begin{equation}\label{G}
C_{res}(k^*)=1-\tilde{\beta} e^{-4k^{*2}R^2},
\end{equation} 
where $\tilde{\beta} = A > 0$ is the annihilation (wide) dip amplitude and $R \ll r_0$ is the dip inverse width.  

Choosing further the fraction of residually correlated particles
    as $\alpha(k^*)=\tilde{\alpha}[1-\lambda(k^*)]$, one may notice
    that the parameters $\tilde{\alpha}$ and $\tilde{\beta}$ enter
    in (\ref{l12}) only through a product $\tilde{\alpha}\tilde{\beta}$,
    the latter can be substituted by a single fit parameter $\beta$.

    In Fig.~\ref{fig4} we present the result of such a fit of the experimental
    $\bar{p}\Lambda \oplus p\bar{\Lambda}$ correlation function.
    The fit quality is quite good ($\chi^2/\mathrm{ndf}=0.87$) and the fitted
    parameters are: $\Re f_0= 0.14\pm 0.66$~fm,
    $\Im f_0= 1.53\pm 1.31$~fm, $\beta=0.034\pm 0.005$
    and $R= 0.48\pm 0.05$~fm. Unfortunately, due to the decoupling of the form
    of the residual correlation function from the scattering
    parameters, the statistics now apperars to be insufficient
    for their reliable determination.

\section{Conclusions}

Study of baryon and antibaryon correlations provides a powerful tool for probing
space-time evolution of heavy ion collisions and for extracting the parameters of strong
interaction between emitted particles.

We reproduced the $p-\Lambda \oplus \bar{p}-\bar{\Lambda}$ and $\bar{p}-\Lambda \oplus p-\bar{\Lambda}$
correlation functions, measured in 10\% most central 
Au+Au collisions by STAR at $\sqrt{s_{NN}}=200$~GeV, using Lednicky and Lyuboshitz 
analytical formalism with the average source radii extracted from the hydrokinetic model (HKM).
To take into account the residual correlations influencing baryon-antibaryon femtoscopic
effects, a modified analytical approximation has been applied. 
The values of the $p\Lambda$ and $p \bar{\Lambda}$ source radii calculated in HKM
are similar, in agreement with theoretical expectations,
and consistent with experimental result for $p-\Lambda \oplus \bar{p}-\bar{\Lambda}$. 
The significantly smaller source size obtained by the STAR Collaboration
for $\bar{p}-\Lambda \oplus p-\bar{\Lambda}$ pairs can be explained by neglecting residual 
correlations at the data analysis.  

The real and imaginary parts of the spin averaged scattering lenghts 
have been extracted 
for baryon-antibaryon pairs when residual correlations are taken into account. We analyse the 
different forms of effective corrections for the residual baryon-antibaryon 
correlations, and find that the simple Gaussian form results in the best fit quality.

The hydrokinetic model including a detailed description of particle correlations 
allows for a precise study of heavy ion collisions.
New high statistics data from RHIC and LHC will provide
measurements of various particle pairs, including baryon-antibaryon ones,
allowing to investigate the particle interactions in these pairs. 
A consistent approach for a wide class of observables will help to understand complex
and unknown features of the evolution of heavy ion collisions. 

\section{Acknowledgments}
The authors are grateful to Iurii Karpenko for his assistance with computer code.
The research was carried out within the scope of the EUREA:
European Ultra Relativistic Energies Agreement (European
Research Group: ``Heavy ions at ultrarelativistic energies'')
and is supported by the National Academy of Sciences of
Ukraine (Agreements F6-2015 and MVC1-2015).

\end{document}